\documentclass[12pt,thmsa]{article}
\def \eq {\begin{equation}}
\def \fim-eq {\end{equation}}
\begin{document}

\author{E. S. Guerra \\
%EndAName
Departamento de F\'{\i}sica \\
Universidade Federal Rural do Rio de Janeiro \\
Cx. Postal 23851, 23890-000 Serop\'edica, RJ, Brazil \\
email: emerson@ufrrj.br\\
}
\title{ TELEPORTATION OF CAVITY FIELD STATES VIA CAVITY QED}
\maketitle

\begin{abstract}
\noindent In this article we discuss two schemes of teleportation of cavity
field states. In the first scheme we consider cavities prepared in a
coherent state and in the second scheme we consider cavities prepared in a
superposition of zero and one Fock states.

\ \newline

\end{abstract}

\section{\protect\bigskip INTRODUCTION\protect\bigskip}

Quantum information and quantum computation opened a completely new prospect
in information processing and are important and active fields of research 
\cite{Nielsen, MathQC, PhysQI}. Teleportation, proposed by Bennett \textit{%
et al }\cite{Bennett}, has important applications in quantum information and
quantum computation \cite{Nielsen}. \ The two ingredients which are
essential in teleportation are the superposition principle and entanglement
and its consequences non-locality. In teleportation a party, Alice, wants to
transfer the unknown quantum state of a given system which someone gives to
her, to a system with another party, Bob, which in principle, is far apart
from Alice. In order to do that, Alice and Bob share a Bell state\ \cite%
{Nielsen, PhysQI}\ (or EPR state \cite{EPR}) in which half of the Bell pair
is with Alice and the other half is with Bob and follow a given prescription
communicating classically with each other. In the end of the process Bob
gets a state identical to the state of the original state in which Alice%
%TCIMACRO{\U{b4}}%
%BeginExpansion
\'{}%
%EndExpansion
s system was prepared and the state of Alice%
%TCIMACRO{\U{b4}}%
%BeginExpansion
\'{}%
%EndExpansion
s system is destroyed since according to the no-cloning theorem \cite%
{Nocloning, Nielsen} it is not possible to clone an arbitrary quantum state.

There has been a lot of theoretical proposals of schemes of teleportation.
In Ref.\ \cite{DavZag} it is proposed a teleportation scheme based on cavity
QED for the teleportation of an atomic state where the cavities are prepared
in a entangled state of zero and one Fock states. In Ref. \cite%
{GuerraTelCascatCS} it is proposed a scheme to teleport an atomic state for
atoms in a cascade configuration making use of cavities prepared in a
coherent state. In Ref. \cite{GuerraTelLambatCS} it is proposed a scheme to
teleport an atomic state for atoms in a lambda configuration making use of
cavities prepared in a coherent state. In Ref. \cite{SongLuZhangGuo} it is
presented a scheme of teleportation where a superposition of zero and one
Fock states is teleported via cavity QED using atoms in a lambda
configuration. An interesting proposition of generating EPR states and
realization of teleportation using a dispersive atom-field interaction is
presented in \cite{ZengGuo}. Teleportation has already been realized
experimentally. It has been demonstrated using optical systems \cite{ExpTel1}
and NMR \cite{ExpTel2}.

In this article we consider Rydberg atoms \cite{Rydat} interacting with a
superconducting cavity \cite{Haroche, Walther}. We study two schemes of
performing teleportation. In the first scheme we consider cavities prepared
in a coherent state and in the second scheme we consider cavities prepared
in a superposition of a zero and a one Fock states.

\section{\protect\bigskip SCHEME 1}

Consider a three-level cascade atom \ $Ak$ with $\mid e_{k}\rangle ,\mid
f_{k}\rangle $ and $\mid g_{k}\rangle $ being the upper, intermediate and
lower atomic state respectively (see Fig. 1). We assume that the transition $%
\mid f_{k}\rangle \rightleftharpoons \mid e_{k}\rangle $ is far enough from
resonance with the cavity central frequency such that only virtual
transitions occur between these states (only these states interact with the
cavity field). In addition we assume that the transition $\mid e_{k}\rangle
\rightleftharpoons \mid g_{k}\rangle $ is highly detuned from the cavity
frequency so that there will be no coupling with the cavity field. Here we
are going to consider only the effect of levels $\mid f_{k}\rangle $ and $%
\mid g_{k}\rangle .$ We do not consider level $\mid e_{k}\rangle $ since it
will not play any role in our scheme. Therefore, we have effectively a
two-level system involving states $\mid f_{k}\rangle $ and $|g_{k}\rangle $.
Considering levels $\mid f_{k}\rangle $ and $\mid g_{k}\rangle ,$ we can
write an effective time evolution operator 
\begin{equation}
U_{k}(t)=e^{i\varphi a^{\dagger }a}\mid f_{k}\rangle \langle f_{k}\mid
+|g_{k}\rangle \langle g_{k}\mid ,  \label{UCasc}
\end{equation}%
where the second term above was put by hand just in order to take into
account the effect of level $\mid g_{k}\rangle $. In (\ref{UCasc}) $a$ $%
(a^{\dagger })$ is the annihilation (creation) operator for the field in the
cavity, $\varphi =g^{2}\tau /$ $\Delta $, \ $g$ is the coupling constant, $%
\Delta =\omega _{e}-\omega _{f}-\omega $ is the detuning \ where \ $\omega
_{e}$ and $\omega _{f}$ \ are the frequencies of the upper and intermediate
levels respectively and $\omega $ is the cavity field frequency and $\tau $
is the atom-field interaction time. Let us take $\varphi =\pi $. \ Now, let
us assume that we let atom $A1$ to interact with cavity $C$ prepared in a
coherent state. Let us define%
\begin{equation}
\mid \psi _{x},\pm \rangle _{Ak}=\frac{1}{\sqrt{2}}(\mid f_{k}\rangle \pm
\mid g_{k}\rangle ),  \label{Psix+/-Ak}
\end{equation}%
and let us assume that atom $A1$ is prepared in a Ramsey cavity $R1$ in the
state $\mid \psi _{x},+\rangle _{A1}.$ Now, \ we assume that we let atom $A1$
to interact with cavity $C1$ prepared in the coherent state $|-\alpha
\rangle _{1}$. Then, taking into account (\ref{UCasc}), the system $A1-C1$
evolves to%
\begin{equation}
\mid \psi \rangle _{A1-C1}=\frac{1}{\sqrt{2}}(\mid f_{1}\rangle |\alpha
\rangle _{1}+\mid g_{1}\rangle |-\alpha \rangle _{1}).
\end{equation}%
If we define the even and odd coherent states 
\begin{eqnarray}
|+\rangle _{Ck} &=&\frac{1}{\sqrt{N_{k}^{+}}}(|\alpha \rangle _{k}+|-\alpha
\rangle _{k}),  \nonumber \\
|-\rangle _{Ck} &=&\frac{1}{\sqrt{N_{k}^{-}}}(|\alpha \rangle _{k}-|-\alpha
\rangle _{k}),  \label{EOCS}
\end{eqnarray}%
with $N_{k}^{\pm }=2\left( 1\pm e^{-2\mid \alpha \mid ^{2}}\right) \cong 2$
\ and $_{Ck}\langle +\mid -\rangle _{Ck}=0$ \cite{EvenOddCS} we have%
\begin{equation}
\mid \psi \rangle _{A1-C1}=\frac{1}{2}[|+\rangle _{C1}(\mid f_{1}\rangle
+\mid g_{1}\rangle )+|-\rangle _{C1}(\mid f_{1}\rangle -\mid g_{1}\rangle )].
\end{equation}%
Making use of (\ref{Psix+/-Ak}) we can rewrite the above expression as 
\begin{equation}
\mid \psi \rangle _{A1-C1}=\frac{1}{\sqrt{2}}(|+\rangle _{C1}|\psi
_{x},+\rangle _{A1}+|-\rangle _{C1}|\psi _{x},-\rangle _{A1}).
\end{equation}%
Now we let atom $A1$ to fly through another cavity $C2$ prepared \ in the
coherent state $|-\alpha \rangle _{2}$ and we have%
\begin{equation}
\mid \psi \rangle _{A1-C1-C2}=\frac{1}{\sqrt{2}}[|+\rangle _{C1}(|+\rangle
_{C2}|\psi _{x},+\rangle _{A1}+|-\rangle _{C2}|\psi _{x},-\rangle
_{A1})+|-\rangle _{C1}(|+\rangle _{C2}|\psi _{x},-\rangle _{A1}+|-\rangle
_{C2}|\psi _{x},+\rangle _{A1})].
\end{equation}%
If $A1$ enters a Ramsey cavity $R2$ where the atomic states are rotated
according to%
\begin{eqnarray*}
|\psi _{x},+\rangle _{A1} &\longrightarrow &\mid f_{1}\rangle \\
|\psi _{x},-\rangle _{A1} &\longrightarrow &\mid g_{1}\rangle
\end{eqnarray*}%
and we detect $\mid f_{1}\rangle $, we get%
\begin{equation}
\mid \Phi ^{+}\rangle _{C1-C2}=\frac{1}{\sqrt{2}}(|+\rangle _{C1}|+\rangle
_{C2}+|-\rangle _{C1}|-\rangle _{C2}).  \label{PHIC1C2CS+}
\end{equation}%
We can also prepare the states 
\begin{equation}
\mid \Phi ^{-}\rangle _{C1-C2}=\frac{1}{\sqrt{2}}(|+\rangle _{C1}|+\rangle
_{C2}-|-\rangle _{C1}|-\rangle _{C2}),  \label{PHIC1C2CS-}
\end{equation}%
and%
\begin{equation}
\mid \Psi ^{+}\rangle _{C1-C2}=\frac{1}{\sqrt{2}}(|+\rangle _{C1}|-\rangle
_{C2}+|-\rangle _{C1}|+\rangle _{C2}),  \label{PSIC1C2CS+}
\end{equation}%
and%
\begin{equation}
\mid \Psi ^{-}\rangle _{C1-C2}=\frac{1}{\sqrt{2}}(|+\rangle _{C1}|-\rangle
_{C2}-|-\rangle _{C1}|+\rangle _{C2}).  \label{PSIC1C2CS-}
\end{equation}%
The states (\ref{PHIC1C2CS+}), (\ref{PHIC1C2CS-}), (\ref{PSIC1C2CS+}) and (%
\ref{PSIC1C2CS-}) are Bell states and form a Bell basis \cite{Nielsen,
PhysQI}.

Now let us assume that Alice keeps with her cavity $C2$ and Bob cavity $C1$.
Then they separate and let us assume that, later on, Alice decides to
teleport a state 
\begin{equation}
\mid \psi \rangle _{C3}=\zeta |+\rangle _{C3}+\xi |-\rangle _{C3}
\label{PsiTelC3}
\end{equation}%
to Bob. Let us see how we can prepare a state like (\ref{PsiTelC3}). Suppose
we prepare cavity $C3$ initially in a coherent state $|-\alpha \rangle _{3}$%
. Then we prepare a two-level atom $B$, with $\mid f\rangle $ and $\mid
g\rangle $ being the upper and lower state respectively, in a coherent
superposition, \ sending $B$ in the lower state $\mid g\rangle $ through a
first Ramsey cavity $K1$ where the atomic states are rotated according to%
\begin{equation}
K_{1}=\frac{1}{\sqrt{2}}\left[ 
\begin{array}{cc}
c_{g} & c_{f} \\ 
-c_{f} & c_{g}%
\end{array}%
\right] ,
\end{equation}%
and we get%
\begin{equation}
\mid \psi \rangle _{B}=c_{f}\mid f\rangle +c_{g}\mid g\rangle .
\end{equation}%
After that, $B$ flies through cavity $C3$ and, taking into account the time
evolution operator (\ref{UCasc}), after $B$ pass through $C3$ the state of
the system $B-C3$, for $\varphi =\pi $, is given by 
\[
\mid \psi \rangle _{B-C3}=c_{f}\mid f\rangle \mid \alpha \rangle
_{3}+c_{g}\mid g\rangle \mid -\alpha \rangle _{3.}
\]%
Then, we send $B$ through a second Ramsey zone $K2$ \ where the atomic
states are rotated according to 
\begin{equation}
K_{2}=\frac{1}{\sqrt{2}}\left[ 
\begin{array}{cc}
1 & -ie^{i\theta } \\ 
-ie^{-i\theta } & 1%
\end{array}%
\right] ,
\end{equation}%
that is,%
\begin{eqnarray}
&\mid &f\rangle \rightarrow \frac{1}{\sqrt{2}}(\mid f\rangle -ie^{-i\theta
}\mid g\rangle ),  \nonumber \\
&\mid &g\rangle \rightarrow \frac{1}{\sqrt{2}}(-ie^{i\theta }\mid f\rangle
+\mid g\rangle ),
\end{eqnarray}%
and therefore, the state of the system $B-C3$ will be 
\begin{eqnarray*}
|\psi \rangle _{B-C3} &=&\frac{1}{\sqrt{2}}[(c_{f}-ie^{i\theta }c_{g})\mid
+\rangle _{C3}+(c_{f}+ie^{i\theta }c_{g})\mid -\rangle _{C3}]\mid f\rangle 
\\
&&+\frac{1}{\sqrt{2}}[(-ie^{-i\theta }c_{f}+c_{g})\mid +\rangle
_{C3})+(-ie^{-i\theta }c_{f}-c_{g})\mid -\rangle _{C3}]\mid g\rangle ,
\end{eqnarray*}%
Now, in order to obtain a state $|\psi \rangle _{C3}$ in cavity $C3$, we
detect atom $B$ in $\mid f\rangle $ or in $\mid g\rangle $. If we detect $%
\mid f\rangle $ we have \ $\zeta =(c_{f}-ie^{i\theta }c_{g})/\sqrt{2}$ and $%
\xi =(c_{f}+ie^{i\theta }c_{g})/\sqrt{2}$. If we detect $\mid g\rangle $ we
have \ $\zeta =(-ie^{-i\theta }c_{f}+c_{g})/\sqrt{2}$ and $\xi
=(-ie^{-i\theta }c_{f}-c_{g})/\sqrt{2}$ .

Now let us see how Alice can teleprot the state (\ref{PsiTelC3}) to Bob.
First we write \ the state formed by the direct product of the Bell state
and this unknown state $\mid \Phi ^{+}\rangle _{C1-C2}\mid \psi \rangle
_{C3} $, that is,

\begin{eqnarray}
&\mid &\psi \rangle _{C1-C2-C3}=\frac{1}{\sqrt{2}}[\zeta (\mid +\rangle
_{C1}\mid +\rangle _{C2}\mid +\rangle _{C3}+\mid -\rangle _{C1}\mid -\rangle
_{C2}\mid +\rangle _{C3})+  \nonumber \\
\xi ( &\mid &+\rangle _{C1}\mid +\rangle _{C2}\mid -\rangle _{C3}+\mid
-\rangle _{C1}\mid -\rangle _{C2}\mid -\rangle _{C3})].
\end{eqnarray}%
If Alice sends an atom $A2$ through $C2$ and $C3$ prepared initially in the
state $|\psi _{x},+\rangle _{A2}$ in a Ramsey cavity $R3$, we have 
\begin{eqnarray}
&\mid &\psi \rangle _{C1-C2-C3-A2}=\frac{1}{\sqrt{2}}[\zeta (\mid +\rangle
_{C1}\mid +\rangle _{C2}\mid +\rangle _{C3}|\psi _{x},+\rangle _{A2}-\mid
-\rangle _{C1}\mid -\rangle _{C2}\mid +\rangle _{C3}|\psi _{x},-\rangle
_{A2})+  \nonumber \\
\xi (- &\mid &+\rangle _{C1}\mid +\rangle _{C2}\mid -\rangle _{C3}|\psi
_{x},-\rangle _{A2}+\mid -\rangle _{C1}\mid -\rangle _{C2}\mid -\rangle
_{C3}|\psi _{x},+\rangle _{A2})],
\end{eqnarray}%
which can be rewritten as%
\begin{eqnarray}
&\mid &\psi \rangle _{C1-C2-C3-A2}=  \nonumber \\
\frac{1}{\sqrt{2}}[ &\mid &\Phi ^{+}\rangle _{C2-C3}(\zeta \mid +\rangle
_{C1}+\xi \mid -\rangle _{C1})|\psi _{x},+\rangle _{A2}+  \nonumber \\
&\mid &\Phi ^{-}\rangle _{C2-C3}(\zeta \mid +\rangle _{C1}-\xi \mid -\rangle
_{C1})|\psi _{x},+\rangle _{A2}+  \nonumber \\
- &\mid &\Psi ^{+}\rangle _{C2-C3}(\zeta \mid -\rangle _{C1}+\xi \mid
+\rangle _{C1})|\psi _{x},-\rangle _{A2}+  \nonumber \\
- &\mid &\Psi ^{-}\rangle _{C2-C3}(-\zeta \mid -\rangle _{C1}+\xi \mid
+\rangle _{C1})|\psi _{x},-\rangle _{A2}].
\end{eqnarray}%
Now Alice sends $A2$ through a Ramsey cavity $R4$ where the atomic states
are rotated according to%
\begin{eqnarray*}
|\psi _{x},+\rangle _{A2} &\longrightarrow &\mid f_{2}\rangle , \\
|\psi _{x},-\rangle _{A2} &\longrightarrow &\mid g_{2}\rangle ,
\end{eqnarray*}%
and if she detects $\mid f_{2}\rangle $ she gets%
\begin{equation}
\mid \psi \rangle _{C1-C2-C3}=\frac{1}{\sqrt{2}}[\mid \Phi ^{+}\rangle
_{C2-C3}(\zeta \mid +\rangle _{C1}+\xi \mid -\rangle _{C1})+\mid \Phi
^{-}\rangle _{C2-C3}(\zeta \mid +\rangle _{C1}-\xi \mid -\rangle _{C1})], 
\nonumber
\end{equation}%
and if she detects $\mid g_{2}\rangle $ she gets%
\begin{equation}
\mid \psi \rangle _{C1-C2-C3}=\frac{1}{\sqrt{2}}[\mid \Psi ^{+}\rangle
_{C2-C3}(\zeta \mid -\rangle _{C1}+\xi \mid +\rangle _{C1})+\mid \Psi
^{-}\rangle _{C2-C3}(-\zeta \mid -\rangle _{C1}+\xi \mid +\rangle _{C1})].
\end{equation}

Notice that%
\begin{eqnarray}
&\mid &\Phi ^{+}\rangle _{C2-C3}=\frac{1}{\sqrt{2}}(|\alpha \rangle
_{2}|\alpha \rangle _{3}+|-\alpha \rangle _{2}|-\alpha \rangle _{3}), 
\nonumber \\
&\mid &\Phi ^{-}\rangle _{C2-C3}=\frac{1}{\sqrt{2}}(|\alpha \rangle
_{2}|-\alpha \rangle _{3}+|-\alpha \rangle _{2}|\alpha \rangle _{3}), 
\nonumber \\
&\mid &\Psi ^{+}\rangle _{C2-C3}=\frac{1}{\sqrt{2}}(|\alpha \rangle
_{2}|\alpha \rangle _{3}-|-\alpha \rangle _{2}|-\alpha \rangle _{3}), 
\nonumber \\
&\mid &\Psi ^{-}\rangle _{C2-C3}=\frac{1}{\sqrt{2}}(|-\alpha \rangle
_{2}|\alpha \rangle _{3}-|\alpha \rangle _{2}|-\alpha \rangle _{3}).
\end{eqnarray}

Now Alice injects $|\alpha \rangle _{2}$ or $|-\alpha \rangle _{2}$ in $C2$
and $|\alpha \rangle _{3}$ or $|-\alpha \rangle _{3}$ in $C3.$ Then Alice
sends a two-level atom $A3$ resonant with the cavity $C2$, with $%
|b_{3}\rangle $ and $|a_{3}\rangle $ being the lower and upper levels
respectively, through $C2$ and a two-level atom $A4$ resonant with the
cavity $C3$, with $|b_{4}\rangle $ and $|a_{4}\rangle $ being the lower and
upper levels respectively, through $C3$. If $Aj$ is sent in the lower state $%
|b_{j}\rangle $, under the Jaynes-Cummings dynamics \cite{Orszag, JCM} we
know that the state $|b_{j}\rangle |0\rangle _{k}$ ($j=3,4$ and $k=2,3$)
does not evolve, however, the state $|b_{j}\rangle |\pm 2\alpha \rangle _{k}$
evolves to $|a_{j}\rangle |\chi _{a}^{\pm }\rangle _{k}+|b_{j}\rangle |\chi
_{b}^{\pm }\rangle _{k}$, where $|\chi _{b}^{\pm }\rangle
_{k}=\sum\limits_{n}C_{n}^{\pm }\cos (gt\sqrt{n})|n\rangle _{k}$ and $|\chi
_{a}^{\pm }\rangle _{k}=-i\sum\limits_{n}C_{n+1}^{\pm }\sin (gt\sqrt{n+1}%
)|n\rangle _{k}$ and $C_{n}^{\pm }=e^{-\frac{1}{2}|\pm 2\alpha
_{k}|^{2}}(\pm 2\alpha _{k})^{n}/\sqrt{n!}$. Therefore, the injection of $%
|\alpha \rangle _{2}$ in $C2$ and $|\alpha \rangle _{3}$ in $C3$ or the
injection of $|-\alpha \rangle _{2}$ in $C2$ and $|-\alpha \rangle _{3}$ in $%
C3$ and the detection of $|a_{3}\rangle $ and $|a_{4}\rangle $ corresponds
to the detection of $\mid \Phi ^{+}\rangle _{C2-C3}$ or $\mid \Psi
^{+}\rangle _{C2-C3}.$ The injection of $|\alpha \rangle _{2}$ in $C2$ and $%
|-\alpha \rangle _{3}$ in $C3$ or the injection of $|-\alpha \rangle _{2}$
in $C2$ and $|\alpha \rangle _{3}$ in $C3$ and the detection of $%
|a_{3}\rangle $ and $|a_{4}\rangle $ corresponds to the detection of $\mid
\Phi ^{-}\rangle _{C2-C3}$ or $\mid \Psi ^{-}\rangle _{C2-C3}$. Therefore,
if Alice detects $\mid f_{2}\rangle $ and injects $|\alpha \rangle _{2}$ in $%
C2$ and $|\alpha \rangle _{3}$ in $C3$ or $|-\alpha \rangle _{2}$ in $C2$
and $|-\alpha \rangle _{3}$ in $C3$ \ and detects $|a_{3}\rangle $ and $%
|a_{4}\rangle ,$ Bob gets 
\begin{equation}
\mid \psi \rangle _{C1}=\zeta \mid +\rangle _{C1}+\xi \mid -\rangle _{C1}.
\label{PsiC1tel1}
\end{equation}%
If Alice detects $\mid f_{2}\rangle $ and injects $|\alpha \rangle _{2}$ in $%
C2$ and $|-\alpha \rangle _{3}$ in $C3$ or $|-\alpha \rangle _{2}$ in $C2$
and $|\alpha \rangle _{3}$ in $C3$ \ and detects $|a_{3}\rangle $ and $%
|a_{4}\rangle ,$ Bob gets 
\begin{equation}
\mid \psi \rangle _{C1}=\zeta \mid +\rangle _{C1}-\xi \mid -\rangle _{C1}.
\label{PsiC1tel2}
\end{equation}%
If Alice detects $\mid g_{2}\rangle $ and injects $|\alpha \rangle _{2}$ in $%
C2$ and $|\alpha \rangle _{3}$ in $C3$ or $|-\alpha \rangle _{2}$ in $C2$
and $|-\alpha \rangle _{3}$ in $C3$ \ and detects $|a_{3}\rangle $ and $%
|a_{4}\rangle ,$ Bob gets 
\begin{equation}
\mid \psi \rangle _{C1}=\zeta \mid -\rangle _{C1}+\xi \mid +\rangle _{C1}.
\label{PsiC1tel3}
\end{equation}%
If Alice detects $\mid g_{2}\rangle $ and injects $|\alpha \rangle _{2}$ in $%
C3$ and $|-\alpha \rangle _{3}$ in $C3$ or $|-\alpha \rangle _{2}$ in $C2$
and $|\alpha \rangle _{3}$ in $C3$ \ and detects $|a_{3}\rangle $ and $%
|a_{4}\rangle ,$ Bob gets%
\begin{equation}
\mid \psi \rangle _{C1}=-\zeta \mid -\rangle _{C1}+\xi \mid +\rangle _{C1}.
\label{PsiC1tel4}
\end{equation}

Notice that in the case of (\ref{PsiC1tel1}) Bob gets the right state and he
has to do nothing else. In the case of (\ref{PsiC1tel2}) Bob can prepare an
atom $A5$ in a Ramsey cavity $R5$ in the state 
\begin{equation}
\mid \psi \rangle _{A5}=\frac{1}{\sqrt{2}}(\mid f_{5}\rangle +\mid
g_{5}\rangle ).
\end{equation}%
and send $A5$ through $C1$. After $A5$ fly through $C1$ Bob gets 
\begin{equation}
\mid \psi \rangle _{C1-A5}=\frac{1}{\sqrt{2}}[\zeta (\mid f_{5}\rangle +\mid
g_{5}\rangle )\mid +\rangle _{C1}-\xi (-\mid f_{5}\rangle +\mid g_{5}\rangle
)\mid -\rangle _{C1}],
\end{equation}%
and if he detects $\mid f_{5}\rangle $ he gets the right state (\ref%
{PsiC1tel1}). In the case of (\ref{PsiC1tel3}) and (\ref{PsiC1tel4}) it is
not possible to fix the states and Bob cannot do anything to get the correct
teleported state. In Fig. 2 we present the setup of the above teleportation
experiment.

\section{SCHEME 2}

We start assuming that we have a cavity $Ck$ prepared in the state 
\begin{equation}
|+\rangle _{Ck}=\frac{(|0\rangle _{k}+|1\rangle _{k})}{\sqrt{2}}.
\label{Ck+}
\end{equation}%
In order to prepare this state, we send a two-level atom $A0$, with $%
|f_{0}\rangle $ and $|e_{0}\rangle $ being the lower and upper level
respectively, in the state 
\begin{equation}
\mid \psi \rangle _{A0}=\frac{1}{\sqrt{2}}(i\mid e_{0}\rangle +\mid
f_{0}\rangle ),
\end{equation}%
through $Ck$, for $A0$ resonant with the cavity. If $g$ is the coupling
constant and $\tau $ the atom-field interaction time, under the
Jaynes-Cummings dynamics, for $g\tau =\pi /2$, \ we know that the state $%
|f_{0}\rangle |0\rangle _{k}$ does not evolve, however, the state $%
|e_{0}\rangle |0\rangle _{k}$ evolves to $-i|f_{0}\rangle |1\rangle _{k}$.
Then, for the cavity initially in the vacuum state $|0\rangle _{k}$, we have%
\begin{equation}
\frac{(|f_{0}\rangle +i|e_{0}\rangle )}{\sqrt{2}}|0\rangle
_{k}\longrightarrow |f_{0}\rangle \frac{(|0\rangle _{k}+|1\rangle _{k})}{%
\sqrt{2}}=|f_{0}\rangle |+\rangle _{Ck}
\end{equation}%
If we start with%
\begin{equation}
\mid \psi \rangle _{A0}=\frac{1}{\sqrt{2}}(-i\mid e_{0}\rangle +\mid
f_{0}\rangle ),
\end{equation}%
we get%
\begin{equation}
|-\rangle _{Ck}=\frac{(|0\rangle _{k}-|1\rangle _{k})}{\sqrt{2}}.
\label{Ck-}
\end{equation}

Now let us assume that cavities $C1$ and $C2$ are prepared in the state (\ref%
{Ck+}). Consider \ an atom $A1$ \ prepared in the state $\mid \psi
_{x},+\rangle _{A1}$ (see (\ref{Psix+/-Ak})) in a Ramsey cavity $R1$. Taking
into account (\ref{UCasc}), after atom $A1$ has passed through the cavities,
we get 
\begin{equation}
\mid \psi \rangle _{A1-C1-C2}=\frac{1}{\sqrt{2}}(|-\rangle _{C1}|-\rangle
_{C2}\mid f_{1}\rangle +|+\rangle _{C1}|+\rangle _{C2}\mid g_{1}\rangle ),
\end{equation}%
Now, if atom $A1$ enters a second Ramsey cavity $R2$ where the atomic states
are rotated according to 
\begin{eqnarray}
&\mid &f_{1}\rangle \rightarrow \frac{1}{\sqrt{2}}(\mid f_{1}\rangle +\mid
g_{1}\rangle ),  \nonumber \\
&\mid &g_{1}\rangle \rightarrow \frac{1}{\sqrt{2}}(-\mid f_{1}\rangle +\mid
g_{1}\rangle ),
\end{eqnarray}%
after we detect $\mid g_{1}\rangle $ we have 
\begin{equation}
\mid \Phi ^{+}\rangle _{C1-C2}=\frac{1}{\sqrt{2}}(|+\rangle _{C1}|+\rangle
_{C2}+|-\rangle _{C1}|-\rangle _{C2}),  \label{PHIC1C2+}
\end{equation}%
It is also easy to prepare%
\begin{equation}
\mid \Phi ^{-}\rangle _{C1-C2}=\frac{1}{\sqrt{2}}(|+\rangle _{C1}|+\rangle
_{C2}-|-\rangle _{C1}|-\rangle _{C2}),  \label{PHIC1C2-}
\end{equation}%
and%
\begin{equation}
\mid \Psi ^{+}\rangle _{C1-C2}=\frac{1}{\sqrt{2}}(|+\rangle _{C1}|-\rangle
_{C2}+|-\rangle _{C1}|+\rangle _{C2}),  \label{PSIC1C2+}
\end{equation}%
and%
\begin{equation}
\mid \Psi ^{-}\rangle _{C1-C2}=\frac{1}{\sqrt{2}}(|+\rangle _{C1}|-\rangle
_{C2}-|-\rangle _{C1}|+\rangle _{C2}).  \label{PSIC1C2-}
\end{equation}%
The states (\ref{PHIC1C2+}), (\ref{PHIC1C2-}), (\ref{PSIC1C2+}) and (\ref%
{PSIC1C2-}) are Bell states and form a Bell basis \cite{Nielsen, PhysQI}.

Now let us assume that Alice keeps with her cavity $C2$ and Bob cavity $C1$.
Then they separate and let us assume that, later on, Alice decides to
teleport a state 
\begin{equation}
\mid \psi \rangle _{C3}=\zeta |+\rangle _{C3}+\xi |-\rangle _{C3}
\label{PsiTelC3-01}
\end{equation}%
to Bob. Now, let us see how we can prepare the state (\ref{PsiTelC3-01}). \
First we send a two-level atom $B$, with $|f\rangle $ and $|e\rangle $ being
the lower and upper level respectively, through a Ramsey cavity $K1$ in the
lower state $|f\rangle $ where the atomic states are rotated according to%
\begin{equation}
K_{1}=\frac{1}{\sqrt{2}}\left[ 
\begin{array}{cc}
c_{f} & c_{e} \\ 
-c_{e} & c_{f}%
\end{array}%
\right] ,
\end{equation}%
and we get%
\begin{equation}
\mid \psi \rangle _{B}=c_{e}\mid e\rangle +c_{f}\mid f\rangle .
\end{equation}%
Next we send $B$ through $C3$ prepared in the vacuum state $|0\rangle _{3}.$
If $g$ is the coupling constant and $\tau $ the atom-field interaction time,
under the Jaynes-Cummings dynamics, for $g\tau =\pi /2$, \ we know that the
state $|f\rangle |0\rangle _{3}$ does not evolve, however, the state $%
|e\rangle |0\rangle _{3}$ evolves to $-i|f\rangle |1\rangle _{3}$. Then we
have%
\begin{equation}
\mid \psi \rangle _{B-C3}=(c_{f}|0\rangle _{3}-ic_{e}|1\rangle _{3})\mid
f\rangle .
\end{equation}%
Therefore, making use of (\ref{Ck+}) and (\ref{Ck-}) the above state can be
written as (\ref{PsiTelC3-01}) with $\zeta =(c_{f}-ic_{e})/\sqrt{2}$ and $%
\xi =(c_{f}+ic_{e})/\sqrt{2}.$

Now let us see how Alice can teleprot the state (\ref{PsiTelC3-01}) to Bob.
First we write \ the state formed by the direct product of the Bell state
and this unknown state $\mid \Phi ^{+}\rangle _{C1-C2}\mid \psi \rangle
_{C3} $, that is,%
\begin{eqnarray}
&\mid &\psi \rangle _{C1-C2-C3}=\frac{1}{\sqrt{2}}[\zeta (\mid +\rangle
_{C1}\mid +\rangle _{C2}\mid +\rangle _{C3}+\mid -\rangle _{C1}\mid -\rangle
_{C2}\mid +\rangle _{C3})+  \nonumber \\
\xi ( &\mid &+\rangle _{C1}\mid +\rangle _{C2}\mid -\rangle _{C3}+\mid
-\rangle _{C1}\mid -\rangle _{C2}\mid -\rangle _{C3})].
\end{eqnarray}%
Now Alice prepares \ an atom $A2$ in the state $\mid \psi _{x},+\rangle
_{A2} $ in a Ramsey cavity $R3$ and send it through cavities $C2$ and $C3$
and we have%
\begin{eqnarray}
&\mid &\psi \rangle _{C1-C2-C3-A2}=  \nonumber \\
\frac{1}{\sqrt{2}}\{\zeta \lbrack &\mid &+\rangle _{C1}(\mid -\rangle
_{C2}\mid -\rangle _{C3}\mid f_{2}\rangle +\mid +\rangle _{C2}\mid +\rangle
_{C3}\mid g_{2}\rangle )+  \nonumber \\
&\mid &-\rangle _{C1}(\mid +\rangle _{C2}\mid -\rangle _{C3}\mid
f_{2}\rangle +\mid -\rangle _{C2}\mid +\rangle _{C3}\mid g_{2}\rangle )]+ \\
\xi \lbrack &\mid &+\rangle _{C1}(\mid -\rangle _{C2}\mid +\rangle _{C3}\mid
f_{2}\rangle +\mid +\rangle _{C2}\mid -\rangle _{C3}\mid g_{2}\rangle )+ 
\nonumber \\
&\mid &-\rangle _{C1}(\mid +\rangle _{C2}\mid +\rangle _{C3}\mid
f_{2}\rangle +\mid -\rangle _{C2}\mid -\rangle _{C3}\mid g_{2}\rangle )]\}
\end{eqnarray}%
Then Alice sends $A2$ through a Ramsey cavity $R4$ where the atomic states
are rotated according to 
\begin{eqnarray}
&\mid &f_{2}\rangle \rightarrow \frac{1}{\sqrt{2}}(\mid f_{2}\rangle +\mid
g_{2}\rangle ),  \nonumber \\
&\mid &g_{2}\rangle \rightarrow \frac{1}{\sqrt{2}}(-\mid f_{2}\rangle +\mid
g_{2}\rangle ),
\end{eqnarray}%
and we have 
\begin{eqnarray}
&\mid &\psi \rangle _{C1-C2-C3-A2}=\frac{1}{2}\{\zeta \lbrack \mid +\rangle
_{C1}(\mid \Phi ^{+}\rangle _{C2-C3}\mid g_{2}\rangle -\mid \Phi ^{-}\rangle
_{C2-C3}\mid f_{2}\rangle )+  \nonumber \\
&\mid &-\rangle _{C1}(\mid \Psi ^{+}\rangle _{C2-C3}\mid g_{2}\rangle +\mid
\Psi ^{-}\rangle _{C2-C3}\mid f_{2}\rangle )]+  \nonumber \\
\xi \lbrack &\mid &+\rangle _{C1}(\mid \Psi ^{+}\rangle _{C2-C3}\mid
g_{2}\rangle -\mid \Psi ^{-}\rangle _{C2-C3}\mid f_{2}\rangle )+  \nonumber
\\
&\mid &-\rangle _{C1}(\mid \Phi ^{+}\rangle _{C2-C3}\mid g_{2}\rangle +\mid
\Phi ^{-}\rangle _{C2-C3}\mid f_{2}\rangle )]\}.
\end{eqnarray}%
If Alice detects $\mid g_{2}\rangle $ she gets 
\begin{eqnarray}
&\mid &\psi \rangle _{C1-C2-C3-A2}=\frac{1}{2}[\zeta (\mid +\rangle
_{C1}\mid \Phi ^{+}\rangle _{C2-C3}+\mid -\rangle _{C1}\mid \Psi ^{+}\rangle
_{C2-C3})  \nonumber \\
+\xi ( &\mid &+\rangle _{C1}\mid \Psi ^{+}\rangle _{C2-C3}+\mid -\rangle
_{C1}\mid \Phi ^{+}\rangle _{C2-C3})],
\end{eqnarray}%
and if she detects $\mid f_{2}\rangle $ she gets%
\begin{eqnarray}
&\mid &\psi \rangle _{C1-C2-C3-A2}=\frac{1}{2}[\zeta (-\mid +\rangle
_{C1}\mid \Phi ^{-}\rangle _{C2-C3}+\mid -\rangle _{C1}\mid \Psi ^{-}\rangle
_{C2-C3})+  \nonumber \\
\xi (- &\mid &+\rangle _{C1}\mid \Psi ^{-}\rangle _{C2-C3}+\mid -\rangle
_{C1}\mid \Phi ^{-}\rangle _{C2-C3})].
\end{eqnarray}%
Notice that 
\begin{eqnarray}
&\mid &\Phi ^{+}\rangle _{C2-C3}=\frac{1}{\sqrt{2}}(|0\rangle _{C2}|0\rangle
_{C3}+|1\rangle _{C2}|1\rangle _{C3}), \\
&\mid &\Phi ^{-}\rangle _{C2-C3}=\frac{1}{\sqrt{2}}(|1\rangle _{C2}|0\rangle
_{C3}+|0\rangle _{C2}|1\rangle _{C3}), \\
&\mid &\Psi ^{+}\rangle _{C2-C3}=\frac{1}{\sqrt{2}}(|0\rangle _{C2}|0\rangle
_{C3}-|1\rangle _{C2}|1\rangle _{C3}), \\
&\mid &\Psi ^{-}\rangle _{C2-C3}=\frac{1}{\sqrt{2}}(|1\rangle _{C2}|0\rangle
_{C3}-|0\rangle _{C2}|1\rangle _{C3}).
\end{eqnarray}%
Now Alice sends a two-level atom $A3$ through $C2$ and a two-level atom $A4$
through $C3$, both resonant with the respective cavity.\ \ Let $%
|f_{3}\rangle $ and $|e_{3}\rangle $ be the lower and upper level of $A3$
respectively and $|f_{4}\rangle $ and $|e_{4}\rangle $ be the lower and
upper level of $A4$ respectively. If $g$ is the coupling constant and $\tau $
the atom-field interaction time, under the Jaynes-Cummings dynamics, for $%
g\tau =\pi /2$, we have 
\begin{eqnarray}
|f_{3}\rangle |f_{4}\rangle &\mid &\Phi ^{+}\rangle _{C2-C3}\longrightarrow 
\frac{1}{\sqrt{2}}(|f_{3}\rangle |f_{4}\rangle -|e_{3}\rangle |e_{4}\rangle
)|0\rangle _{C2}|0\rangle _{C3}=|\Phi ^{-}\rangle _{A3-A4}|0\rangle
_{C2}|0\rangle _{C3},  \label{PHIA3A4-} \\
|f_{3}\rangle |f_{4}\rangle &\mid &\Phi ^{-}\rangle _{C2-C3}\longrightarrow -%
\frac{i}{\sqrt{2}}(|f_{3}\rangle |e_{4}\rangle +|e_{3}\rangle |f_{4}\rangle
)|0\rangle _{C2}|0\rangle _{C3}=|\Psi ^{+}\rangle _{A3-A4}|0\rangle
_{C2}|0\rangle _{C3},  \label{PSIA3A4+} \\
|f_{3}\rangle |f_{4}\rangle &\mid &\Psi ^{+}\rangle _{C2-C3}\longrightarrow 
\frac{1}{\sqrt{2}}(|f_{3}\rangle |f_{4}\rangle +|e_{3}\rangle |e_{4}\rangle
)|0\rangle _{C2}|0\rangle _{C3}=|\Phi ^{+}\rangle _{A3-A4}|0\rangle
_{C2}|0\rangle _{C3},  \label{PHIA3A4+} \\
|f_{3}\rangle |f_{4}\rangle &\mid &\Psi ^{-}\rangle _{C2-C3}\longrightarrow 
\frac{i}{\sqrt{2}}(|f_{3}\rangle |e_{4}\rangle -|e_{3}\rangle |f_{4}\rangle
)|0\rangle _{C2}|0\rangle _{C3}=|\Psi ^{-}\rangle _{A3-A4}|0\rangle
_{C2}|0\rangle _{C3}.  \label{PSIA3A4-}
\end{eqnarray}%
Then, in the case that Alice had detected $\mid g_{2}\rangle ,$ we have 
\begin{eqnarray}
&\mid &\psi \rangle _{C1-C2-C3-A2}=\frac{1}{2}[\zeta (\mid +\rangle
_{C1}|\Phi ^{-}\rangle _{A3-A4}+\mid -\rangle _{C1}|\Phi ^{+}\rangle
_{A3-A4})  \nonumber \\
+\xi ( &\mid &+\rangle _{C1}|\Phi ^{+}\rangle _{A3-A4}+\mid -\rangle
_{C1}|\Phi ^{-}\rangle _{A3-A4})],
\end{eqnarray}%
and if she had detected $\mid f_{2}\rangle ,$ we have%
\begin{eqnarray}
&\mid &\psi \rangle _{C1-C2-C3-A2}=\frac{1}{2}[\zeta (-\mid +\rangle
_{C1}|\Psi ^{+}\rangle _{A3-A4}+\mid -\rangle _{C1}|\Psi ^{-}\rangle
_{A3-A4})+  \nonumber \\
\xi (- &\mid &+\rangle _{C1}|\Psi ^{-}\rangle _{A3-A4}+\mid -\rangle
_{C1}|\Psi ^{+}\rangle _{A3-A4})].
\end{eqnarray}%
Now we define 
\begin{equation}
\Sigma _{x}=\sigma _{x}^{3}\sigma _{x}^{4},
\end{equation}%
where%
\begin{equation}
\sigma _{x}^{k}=\mid f_{k}\rangle \langle e_{k}\mid +\mid e_{k}\rangle
\langle f_{k}\mid ,
\end{equation}%
and we have%
\begin{eqnarray}
\Sigma _{x} &\mid &\Psi ^{\pm }\rangle _{A3-A4}=\pm \mid \Psi ^{\pm }\rangle
_{A3-A4},  \nonumber \\
\Sigma _{x} &\mid &\Phi ^{\pm }\rangle _{A3-A4}=\pm \mid \Phi ^{\pm }\rangle
_{A3-A4}.  \label{AVSIGMAx}
\end{eqnarray}%
Therefore, we can distinguish between $(\mid \Psi ^{+}\rangle _{A3-A4},\mid
\Phi ^{+}\rangle _{A3-A4})$ and $(\mid \Psi ^{-}\rangle _{A3-A4},\mid \Phi
^{-}\rangle _{A3-A4})$ performing measurements of $\Sigma _{x}=\sigma
_{x}^{3}\sigma _{x}^{4}$. In order to do so, we proceed as follows. We make
use of

\begin{equation}
K_{k}=\frac{1}{\sqrt{2}}\left[ 
\begin{array}{cc}
1 & -1 \\ 
1 & 1%
\end{array}%
\right] ,
\end{equation}%
or%
\begin{equation}
K_{k}=\frac{1}{\sqrt{2}}(\mid f_{k}\rangle \langle f_{k}\mid -\mid
f_{k}\rangle \langle e_{k}\mid +\mid e_{k}\rangle \langle f_{k}\mid +\mid
e_{k}\rangle \langle e_{k}\mid ),  \label{KkEPR}
\end{equation}%
to gradually unravel the Bell states. The eigenvectors of the operators $%
\sigma _{x}^{k}$ are%
\begin{equation}
|\psi _{x},\pm \rangle _{Ak}=\frac{1}{\sqrt{2}}(\mid f_{k}\rangle \pm \mid
e_{k}\rangle ),  \label{PSIxEPR}
\end{equation}%
and we can rewrite the Bell states as 
\begin{eqnarray}
&\mid &\Phi ^{\pm }\rangle _{A3-A4}=\frac{1}{2}[|\psi _{x},+\rangle
_{A3}(\mid f_{4}\rangle \pm \mid e_{4}\rangle )+|\psi _{x},-\rangle
_{A3}(\mid f_{4}\rangle \mp \mid e_{4}\rangle )],  \nonumber \\
&\mid &\Psi ^{\pm }\rangle _{A3-A4}=\frac{1}{2}[|\psi _{x},+\rangle
_{A3}(\mid e_{4}\rangle \pm \mid f_{4}\rangle )+|\psi _{x},-\rangle
_{A3}(\mid e_{4}\rangle \mp \mid f_{4}\rangle )].  \label{EPRPSIx}
\end{eqnarray}%
Let us take for instance (\ref{PHIA3A4+}), 
\begin{equation}
\mid \Phi ^{+}\rangle _{A3-A4}=\frac{1}{\sqrt{2}}(\mid f_{3}\rangle \mid
f_{4}\rangle +\mid e_{3}\rangle \mid e_{4}\rangle ).
\end{equation}%
Applying $K_{3}$ to this state we have%
\begin{equation}
K_{3}\mid \Phi ^{+}\rangle _{A3-A4}=\frac{1}{2}[|f_{3}\rangle (\mid
f_{4}\rangle -\mid e_{4}\rangle )+|e_{3}\rangle (\mid f_{4}\rangle +\mid
e_{4}\rangle )].  \label{EPRK1PSI12}
\end{equation}%
Now, we compare (\ref{EPRK1PSI12}) and (\ref{EPRPSIx}). We see that the
rotation by $K_{3}$ followed by the detection of $|e_{3}\rangle $
corresponds to the detection of the the state $|\psi _{x},+\rangle _{A3}$
whose eigenvalue of $\sigma _{x}^{3}$ is $+1$. After we detect $%
|e_{3}\rangle $, we get%
\begin{equation}
\mid \psi \rangle _{A4}=\frac{1}{\sqrt{2}}(\mid f_{4}\rangle +\mid
e_{4}\rangle ),
\end{equation}%
that is, we have got 
\begin{equation}
\mid \psi \rangle _{A4}=|\psi _{x},+\rangle _{A4}.  \label{EPRPSI2x}
\end{equation}%
If we apply (\ref{KkEPR}) for $k=4$ to the state (\ref{EPRPSI2x}) we get%
\begin{equation}
K_{4}\mid \psi \rangle _{A4}=|e_{4}\rangle .
\end{equation}%
We see that the rotation by $K_{4}$ followed by the detection of $%
|e_{4}\rangle $ corresponds to the detection of the the state $|\psi
_{x}^{4},+\rangle $ whose eigenvalue of $\sigma _{x}^{4}$ is $+1$. The same
applies to (\ref{PSIA3A4+}).

Summarizing, we have two possible sequences of atomic state rotations
through $K_{k}$ and detections of $\mid f_{k}\rangle $ or $\mid e_{k}\rangle 
$ and the corresponding states $|\psi _{x}^{k},\pm \rangle $ where $k=3$ and 
$4$ which corresponds to the measurement of the eigenvalue $+1$ of the
operator $\Sigma _{x}$ given by (\ref{AVSIGMAx}) and the detection of (\ref%
{PHIA3A4+}) or (\ref{PSIA3A4+}) corresponds to 
\begin{eqnarray}
(K_{3}, &\mid &e_{3}\rangle )(K_{4},\mid e_{4}\rangle )\longleftrightarrow
|\psi _{x},+\rangle _{A3}|\psi _{x},+\rangle _{A4},  \nonumber \\
(K_{3}, &\mid &f_{3}\rangle )(K_{4},\mid f_{4}\rangle )\longleftrightarrow
|\psi _{x},-\rangle _{A3}|\psi _{x},-\rangle _{A4}.  \label{TestSIGMAx+}
\end{eqnarray}

Considering (\ref{PHIA3A4-}) and (\ref{PSIA3A4-}) we have 
\begin{eqnarray}
(K_{3}, &\mid &e_{3}\rangle )(K_{4},\mid f_{4}\rangle )\longleftrightarrow
|\psi _{x},+\rangle _{A3}|\psi _{x},-\rangle _{A4},  \nonumber \\
(K_{1}, &\mid &f_{3}\rangle )(K_{4},\mid e_{4}\rangle )\longleftrightarrow
|\psi _{x},-\rangle _{A3}|\psi _{x},+\rangle _{A4},  \label{TestSIGMAx-}
\end{eqnarray}%
which corresponds to the measurement of the eigenvalue $-1$ of the operator $%
\Sigma _{x}$ \ given by (\ref{AVSIGMAx}).

Therefore, after the sequence $\mid g_{2}\rangle (K_{3},\mid e_{3}\rangle
)(K_{4},\mid e_{4}\rangle )$ or $\mid g_{2}\rangle (K_{3},\mid f_{3}\rangle
)(K_{4},\mid f_{4}\rangle )$ Bob gets,%
\begin{equation}
\mid \psi \rangle _{C1}=\zeta \mid -\rangle _{C1}+\xi \mid +\rangle _{C1}.
\label{Psitel1}
\end{equation}%
After the sequence $\mid g_{2}\rangle (K_{3},\mid e_{3}\rangle )(K_{4},\mid
f_{4}\rangle )$ or $\mid g_{2}\rangle (K_{3},\mid f_{3}\rangle )(K_{4},\mid
e_{4}\rangle )$ Bob gets,%
\begin{equation}
\mid \psi \rangle _{C1}=\zeta \mid +\rangle _{C1}+\xi \mid -\rangle _{C1}.
\label{Psitel2}
\end{equation}%
After the sequence $\mid f_{2}\rangle (K_{3},\mid e_{3}\rangle )(K_{4},\mid
e_{4}\rangle )$ or $\mid f_{2}\rangle (K_{3},\mid f_{3}\rangle )(K_{4},\mid
f_{4}\rangle )$ Bob gets,%
\begin{equation}
\mid \psi \rangle _{C1}=-\zeta \mid +\rangle _{C1}+\xi \mid -\rangle _{C1}.
\label{Psitel3}
\end{equation}%
After the sequence $\mid f_{2}\rangle (K_{3},\mid e_{3}\rangle )(K_{4},\mid
f_{4}\rangle )$ or $\mid f_{2}\rangle (K_{3},\mid f_{3}\rangle )(K_{4},\mid
e_{4}\rangle )$ Bob gets,%
\begin{equation}
\mid \psi \rangle _{C1}=\zeta \mid -\rangle _{C1}-\xi \mid +\rangle _{C1}.
\label{Psitel4}
\end{equation}

In the case of (\ref{Psitel2}) Bob gets the right state and he has to do
nothing else. In the case (\ref{Psitel1}), consider a two-level atom \ $A5$
with $\mid e_{5}\rangle $ and $\mid f_{5}\rangle $ being the upper and lower
atomic state respectively such that the transition $\mid f_{5}\rangle
\rightleftharpoons \mid e_{5}\rangle $ is far enough from resonance with the
cavity central frequency so that we have a dispersive atom-field
interaction. \ Then the \ time evolution operator is given by 
\begin{equation}
U(t)=e^{-i\varphi (a^{\dagger }a+1)}\mid e\rangle \langle e\mid +e^{i\varphi
a^{\dagger }a}\mid f\rangle \langle f\mid ,  \label{UCascef}
\end{equation}%
where $\varphi =g^{2}\tau /$ $\Delta $. Then, for 
\[
\mid \psi \rangle _{A5}=\frac{1}{\sqrt{2}}(\mid f_{5}\rangle +\mid
e_{5}\rangle ),
\]%
and for $\varphi =\pi $, we have%
\begin{eqnarray*}
&\mid &\psi \rangle _{A5}\mid +\rangle _{C1}\longrightarrow \frac{1}{\sqrt{2}%
}(-\mid f_{5}\rangle +\mid e_{5}\rangle )\mid -\rangle _{C1}, \\
&\mid &\psi \rangle _{A5}\mid -\rangle _{C1}\longrightarrow \frac{1}{\sqrt{2}%
}(-\mid f_{5}\rangle +\mid e_{5}\rangle )\mid +\rangle _{C1},
\end{eqnarray*}%
and after sending atom $A5$ through $C1$ Bob gets the right state. Notice
finally that it is not possible to fix the states (\ref{Psitel3}) and (\ref%
{Psitel4}). In Fig. 3 we present the setup of the above teleportation
experiment.

\section{\protect\bigskip CONCLUSION}

In this article we have studied two schemes of teleportation of cavity field
states by the interaction of Rydberg atoms with superconducting cavities. In
the first scheme we show how to teleport a state which is a superposition of
an even and an odd coherent state. In the second scheme the state to be
teleported is a state constructed with \ zero and one Fock states. In both
schemes it is possible to achieve teleportation only with 50\% of success
since it is not possible to handle cavity field states and to fix all the
wrong teleported states as in the case of atomic states which can be rotated
easily.

\end{document}